\documentclass[10pt,twocolumn,letterpaper]{article}

\usepackage[pagenumbers]{wacv} 

\usepackage{graphicx}
\usepackage[table]{xcolor}
\usepackage{amsmath}
\usepackage{amssymb}
\usepackage{booktabs}
\usepackage{array}
\usepackage{caption}
\usepackage{multirow}
\usepackage{supertabular}
\usepackage{float}
\usepackage{stfloats}
\graphicspath{ {./figures/} }
\usepackage[breaklinks,pagebackref,colorlinks,linkcolor=blue,urlcolor=magenta]{hyperref}

\usepackage[capitalize]{cleveref}
\crefname{section}{Sec.}{Secs.}
\Crefname{section}{Section}{Sections}
\Crefname{table}{Table}{Tables}
\crefname{table}{Tab.}{Tabs.}

\def\wacvPaperID{2679} 
\def\confName{WACV}
\def\confYear{2025}

\begin{document}

\title{Sifting through the haystack 
- efficiently finding rare animal behaviors in large-scale datasets}

\author{
Shir Bar$^{1,2,*}$ \\ 
{\tt\small shirbar@tauex.tau.ac.il} \and
Or Hirschorn$^{3}$ \\
{\tt\small orhirschorn@mail.tau.ac.il} \and
Roi Holzman$^{1,2}$\\
{\tt\small holzman@tau.ac.il} \and
Shai Avidan$^{3}$ \\
{\tt\small avidan@tau.eng.ac.il}\\[0.2cm]
\small $^{1}$ School of Zoology, the Faculty of Life Sciences, Tel Aviv University, Israel,
\small $^{2}$The Interuniversity Institute for Marine Sciences in Eilat, Israel\\
\small $^{3}$ School of Electrical Engineering, the Faculty of Engineering, Tel Aviv University, Israel
$^*$\scriptsize Corresponding author: {\tt shirbar@tauex.tau.ac.il}
}

\maketitle

\begin{abstract}
In the study of animal behavior, researchers often record long continuous videos, accumulating into large-scale datasets. 
However, the behaviors of interest are often rare compared to routine behaviors. 
This incurs a heavy cost on manual annotation,
forcing users to sift through many samples before finding their needles. 
We propose a pipeline to efficiently sample rare behaviors from large datasets, enabling the creation of training datasets for rare behavior classifiers. Our method only needs an unlabeled animal pose or acceleration dataset as input and makes no assumptions regarding the type, number, or characteristics of the rare behaviors.

Our pipeline is based on a recent graph-based anomaly detection model for human behavior, which we apply to this new data domain. 
It leverages anomaly scores to automatically label normal samples while directing human annotation efforts toward anomalies.
In research data, anomalies may come from many different sources (\eg, signal noise versus true rare instances).
Hence, the entire labeling budget is focused on the abnormal classes, letting the user review and label samples according to their needs.

We tested our approach on three datasets of freely-moving animals, acquired in the laboratory and the field.
We found that graph-based models are particularly useful when studying motion-based behaviors in animals, yielding good results while using a small labeling budget.
Our method consistently outperformed traditional random sampling, offering an average improvement of 70\% in performance and creating datasets even when the behavior of interest was only 0.02\% of the data. Even when the performance gain was minor (\eg, when the behavior is not rare), our method still reduced the annotation effort by half \footnote{Our code is available at \url{https://github.com/shir3bar/SiftingTheHaystack}.}.
\end{abstract}

\section{Introduction}

\begin{figure*}[htp]
  \centering
    \includegraphics[scale=0.3]{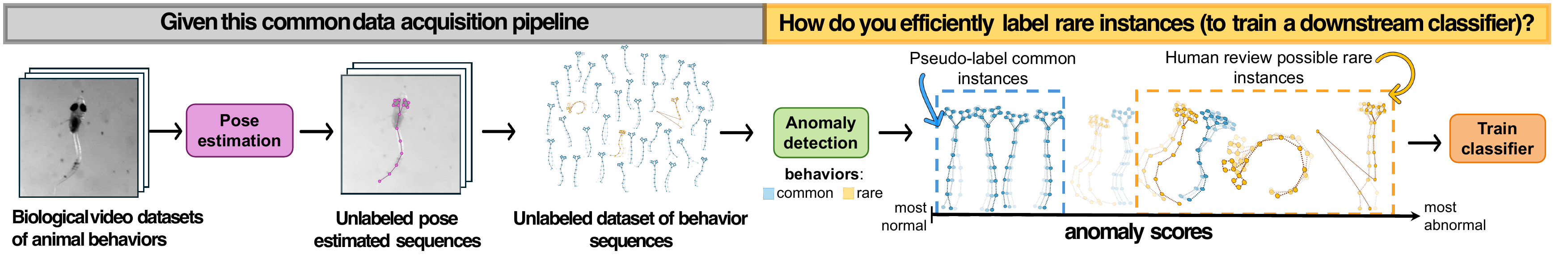}
    \caption{Problem setup and approach. Biologists studying freely behaving animals often accumulate large datasets of behavioral sequences where their behaviors of interest are rare in the data (blue skeletons - common, yellow skeletons - rare). Getting an initial labeled dataset of these behaviors requires a substantial labeling effort.
    Examples of rare instances in larval fish behavior data (yellow skeletons in the yellow dashed box, left to right) include S-starts, C-bends, and noise. The former two are behaviors of interest, the latter is a nuisance feature of large-scale datasets.
    Our method enables the user to efficiently find and annotate rare behaviors with only the unlabeled data as input.
    }
    \label{figure:problem}
\end{figure*}

Rare events are integral to our understanding of animal ecology \cite{cook2021philosophy}. 
With the advance of high-resolution recording devices being deployed continuously by ecologists \cite{tuia2022perspectives} many behaviors, like prey capture in fish \cite{bar2023assessing} or diving behavior in sea birds \cite{tanigaki2024automatic}, become a needle to be searched in an ever-growing haystack.

Our goal is to train classifiers
that can detect these rare behaviors in an unlabeled dataset. To do that, we need to annotate a sufficient number of instances to train a rare behavior classifier. The problem is that finding even a small number of rare events is a (very) time-consuming process to begin with. Thus we consider the setting of using no initial samples of these behaviors. Once equipped with said classifier, we can scan large volumes of unlabeled data and detect rare events of interest.

Many prior works have considered the setting where few or even many rare samples exist for training a preliminary model \cite{mullapudi2021learning}. 
However, we claim a more common use case in biology is where a researcher is confronted with a large dataset of unlabeled recordings with no rare examples relevant to the dataset \cite{pelleg2004active}.
Thus, we ask how, when
given such a dataset,
does one find rare behaviors without any prior assumptions on the type, number, or even the existence of these behaviors in the data (\Cref{figure:problem}).

Unsupervised anomaly detection methods could provide an answer, seeing as these methods are geared towards finding irregular ("anomalous") samples without any annotated examples. 
However, unsupervised methods usually yield lower performance compared to supervised methods, posing a trade-off between performance and annotation effort.
The problem with such methods is that they may find many different types of anomalies (e.g., sensor noise versus abnormal behavior, \Cref{figure:problem}), and not all of these will be of interest to the researcher. 
Hence, supervised methods work well, but require labeled training data that is hard to acquire, and anomaly detection methods work on unlabeled data, but may not have high enough specificity for our needs.
To mitigate the limitations of both approaches and benefit from their strengths, 
we propose a new pipeline to quickly find rare animal behaviors using anomaly detection, label them manually, and then train a semi-supervised classifier (\Cref{figure:problem}).

\begin{figure*}[htp]
  \centering
    \includegraphics[scale=0.8]
    {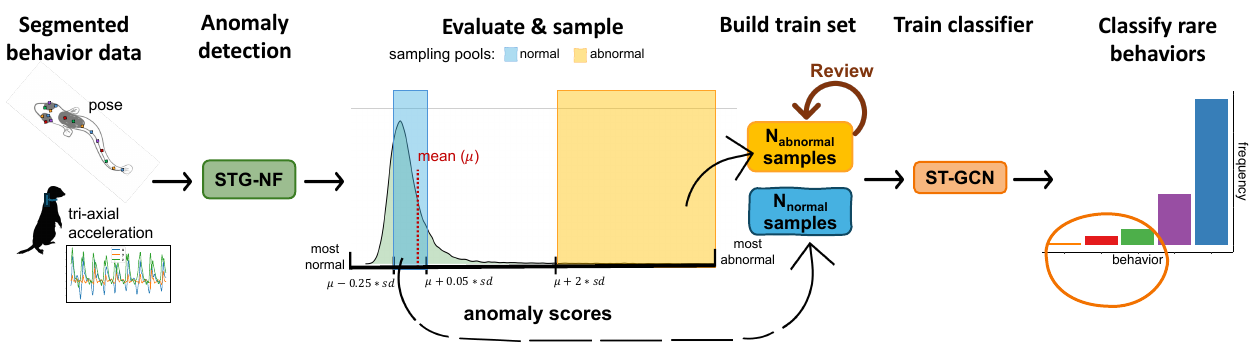}
    \caption{Overview of our proposed framework.
    Starting from behavior data segmented to clips, we first train an anomaly detector on all the available train data. We then evaluate the same dataset and sample the dataset according to the resulting anomaly scores distribution. We draw normal samples from the center of the distribution (blue rectangle) and abnormal samples from the right tail of the distribution (yellow rectangle). We then manually review only the abnormal and train a binary classifier to find rare behaviors.}
    \label{figure:overview}
\end{figure*}

Specifically, we test the application of a recently introduced graph-based human pose anomaly detector \cite{hirschorn2023normalizing} and adapt it to this new data domain. 
For our pipeline, we run the anomaly detector on the data.
This ranks each instance with an anomaly score (a higher score means a more anomalous example). 
We then sample the resulting score distribution in two distinct regions to create a labeled dataset (\Cref{figure:overview}). 
We take samples that are close to the mean anomaly score and label them as normal, without review. 
For the rare behaviors, we sample among the high-scoring instances for anomalous examples.
However, the anomalies may be irrelevant and the chance of false positives is high. 
We therefore let a user manually scan the anomalous examples and annotate the ones that fit their needs. 
Once enough rare samples are found, we can train a supervised classifier.  \Cref{figure:problem,figure:overview} give an overview of the pipeline.

We test our approach on 3 biological datasets of animal behavior from two data modalities that are ubiquitous in behavior research - pose estimation, and accelerometry. In each case, we represent the data using spatio-temporal graphs. 
We find that graphs are an excellent abstraction that works well on biological data, particularly in the context of rare behaviors.
Moreover, it is also very light, leading to fast processing times. 
Finally, we show that, for the same labeling budget,  the effort it takes to label rare examples using our pipeline is considerably lower, 
compared to sequential or random sampling of the data.

\textbf{Our contributions:}
1. Applying a recent anomaly detection framework for rare behavior detection in animals.\\
2. A pipeline that enables the detection of rare animal behaviors in a given dataset without prior assumptions on the type or number of instances of the behaviors in the data.\\
3. Showcasing the power of graph-based representations and Graph Convolutional Networks (GCNs) for behavior analysis in animals.

\section{Related work}
Our work builds on recent progress in the field of anomaly detection in human action recognition relying on GCNs. We combine it into a semi-supervised scheme and train a graph-based classifier to detect rare behaviors.

\paragraph{Anomaly detection in action recognition}
In most traditional anomaly detection settings (\eg,\cite{bergmann2019mvtec, huang2020surface}) the anomalies are stark deviations either 
 in appearance-based statistics or in semantics.
In contrast, most anomaly detection benchmarks for human actions are constructed from surveillance camera footage\cite{abubakar2024systematic}. 
Here, appearance is often difficult to resolve but anomalies are associated with different motions, rather than different appearance or semantics.
The behaviors that are considered anomalies differ between datasets but generally include the use of vehicles, violent behaviors, etc \cite{liu2018future, acsintoae2022ubnormal, sultani2018real}.
Recent work \cite{hirschorn2023normalizing} capitalized on this insight by performing anomaly detection on pose space, rather than pixel space, using Graph Convolutional Networks (GCNs). 
Thus, focusing the model on motion-related anomalies. 
We test their approach for detecting abnormal animal behaviors.

\paragraph{Graph-based analysis in animal behavior}
Working on graph representations has the advantage of abstracting away information that is irrelevant for behavior analysis like camera angle or lighting conditions.
This is beneficial when dealing with rare behaviors, where the model has a limited amount of examples to learn from, as it focuses the model on relevant features of the data.
But, by abstracting appearance we explicitly target kinematic behaviors and cannot detect some classes of behaviors (\eg, camouflage or color change).  
Other applications of graph-based analyses in the animal domain:
\cite{wu2020deepposegraph} improves pose estimation performance by fitting a graph-based model to laboratory animal data, leveraging data from unlabeled frames in a semi-supervised fashion.
\cite{zhu2021graph} counted animals from drones using graph regularized networks.
\cite{li2021coarse} animal pose and shape estimation using GCN-based models for 3D mesh refinement.
\cite{mullen2023poser} uses Spatial-Temporal-GCNs (ST-GCNs, \cite{yan2018spatial}) to classify behaviors of larval fish in an experimental setup.
To the best of our knowledge, we are the first to test a graph-based approach for rare behavior detection in animals.

\paragraph{Animal behavior analysis}
A common pipeline in modern behavior analysis includes extracting pose, then calculating kinematics parameters pre-determined by the researchers to create a time series, then either doing unsupervised clustering or training a tree-based classifier (\eg \cite{chakravarty2019novel,tanigaki2024automatic,johnson2020probabilistic, nilsson2020simple, weinreb2024keypoint}).
Pre-defining the kinematics parameters is not always possible or desirable. When we have no apriori knowledge of the behavior of interest, or when we do not want to introduce human bias into the process of behavior analysis \cite{mccullough2021unsupervised}.
We propose working directly on the acquired pose, dispensing the need to pre-determine which kinematic features are of interest.
Recently, \cite{mullen2023poser} proposed the use of ST-GCNs for supervised multi-class classification of swimming behaviors in fish using pose data directly. 
In our work, we ask how we can quickly find instances of rare behaviors in order to train such classifiers.

\paragraph{Rare events detection in animals}
The ability to observe rare behaviors is thought to be limiting to ecological inference \cite{dixon2005improving}. Nevertheless, in animals, this problem has received relatively little attention, especially in a computer vision context.
Therefore there are few large datasets geared towards testing this problem directly. 
Recent applications include identifying abnormal behavior induced by exposure to chemicals \cite{green2023deep}, and identifying rare behaviors in real-time from bio-loggers deployed on-board sea-birds \cite{tanigaki2024automatic}.
Both applications rely on feature extraction from the raw sensor data followed by tree-based (Random Forest/Isolation Forest) analysis. 
Neither of these works operates directly on sensor data nor tests the effect of similarity between behaviors or behavior rarity.

\paragraph{Semi-supervised and active learning for rare category detection}
Active learning for rare category detection has received a lot of attention over the years (reviewed by \cite{mullapudi2021learning}, but also see \cite{pelleg2004active}).
In active learning, feature representations are iteratively updated according to user feedback. 
We do not consider our work to be active learning as we do not update the anomaly detector ranking the abnormal behaviors.

Mullapudi \etal \cite{mullapudi2021learning} highlight the fact that previous works on rare category learning have many unrealistic assumptions. For example, the assumption you have a large initial dataset of rare behaviors to learn from. 
Their solution, which assumes a small (N=5) set of rare examples, is elegant and forms a basis for further rare category research.

Our method differs from theirs by relaxing two key assumptions that are relevant to biological research datasets.
First, we do not require any rare samples. Finding even 5 instances of a rare behavior in a large dataset will require an intensive labeling effort.
Second, we do not limit the number of rare categories. 
Mullapudi \etal \cite{mullapudi2021learning} assume a single rare category is learned at a time when in biological datasets several types of rare categories may appear. This means the intra-class variability of the rare class will be high in real scenarios.

\section{Approach}
We start by introducing the models used in our pipeline and how we applied them to datasets of animal behaviors.
We then move on to describing our proposed labeling pipeline assuming a limited labeling budget. 
Seeing as truly rare behaviors are often discarded from datasets as outliers, or otherwise not even documented, we perform a rarity experiment where we test the pipeline under increasing rarity levels.

Experimental datasets often have noisy or mislabeled samples (or both), therefore we start by establishing performance on a synthetic dataset. 
Using this dataset we probe the performance of the models and our labeling pipeline. 
We then test our method on three publicly available datasets collected by biologists in the laboratory and the field.

\subsection{Models}

\paragraph{Spatio-Temporal Graph Normalizing Flows (STG-NF) for anomaly detection.}
Normalizing flows \cite{dinh2014nice} are a class of unsupervised models for density estimation that define an invertible mapping between a data space \(\mathcal{X}\) and a latent space \(\mathcal{Z}\), typically assuming a tractable density in \(\mathcal{Z}\), such as a Gaussian distribution. 
This mapping is constructed through a sequence of invertible transformations parameterized by a neural network, ensuring that both the mapping and its Jacobian are computationally efficient. The model allows for direct estimation of the data distribution \(p_X(x)\) by leveraging the change of variables formula, enabling precise computation of the probability density.

We use the STG-NF model \cite{hirschorn2023normalizing}, which uses normalizing flows to directly map graph sequence data to a latent Gaussian distribution. It uses spatio-temporal graph convolutions to model anatomical and motion-based graph connections. Training is done by minimizing the negative log-likelihood of the data probability.
We applied the model to long pose sequences by segmenting them using a sliding window of size {\scriptsize$\mathit{f}=8$}. 
At inference, we obtain sequence-level anomaly scores by taking the mean score across all segments.

\paragraph{Spatial Temporal Graph Convolutional Networks (ST-GCN) for  action classification.}
Yan \etal \cite{yan2018spatial} used graphs to model the dynamics of human skeletons, proposing the spatio-temporal GCN building blocks later used by STG-NF. We used a shallower version of the ST-GCN \cite{yan2018spatial} model, reducing the number of layers in the model. This served to prevent overfitting to the smaller datasets we used.

\subsection{Labeling pipeline} 
To mitigate the trade-off between labeling effort and model performance, we test the feasibility of using anomaly scores produced by the unsupervised STG-NF to train an ST-GCN classifier using a semi-supervised approach.

The basic scheme is depicted in \Cref{figure:overview}. 
For a given dataset, we first train an unsupervised STG-NF model on all available training data. 
We then use the trained model to rank the samples of the \textbf{train data} according to the assigned anomaly score. 
From this ranked dataset we sample instances from two areas in the anomaly score distribution -  the tail (i.e., most abnormal) and around the mean of the distribution (i.e., most representative normal).
We let a human review \textbf{only} the abnormal samples, searching for rare behaviors and labeling the samples as either "normal" (for the common behaviors) or "abnormal" (for the rare behaviors). 
The normal samples obtained this way can be viewed as "hard negatives" as they are closer to positive samples in the feature space. 
In addition to these, we sample instances around the mean of the anomaly score distribution and pseudo-label them as "normal" without human review. 

We sample an equal number of top- and bottom-scoring samples, however, the review process of the top-scoring samples may change the data balance.
We train an ST-GCN classifier using this dataset and evaluate it on the original test set of each dataset. 
We keep this test stable across all experiments, regardless of train dataset size or induced behavioral rarity (see below).

We compare our method to a baseline of random sampling - drawing samples from the training set at random and reviewing them, which is the prevalent method used by many biologists.
We repeat this process using different labeling efforts {\scriptsize $\mathit{N_{reviewed}} = 30, 60, 100, 300, 600, 1000$}. 
We note that the size of the training dataset created by our method is {\scriptsize $\mathit{N_{reviewed}+N_{pseudolabeled}} = 2*\mathit{N_{reviewed}}$} while that of random sampling is $\mathit{N_{reviewed}}$. 
Thus, for the same labeling effort, our method creates a dataset that is double the size.

\section{Evaluation}
\paragraph{Rarity experiments.}
We are mainly interested in very rare behaviors, ones that are so rare that it is difficult to create datasets to investigate the efficacy of models in detecting them.
So, we put our pipeline to the test not only on the observed rarities in each dataset but also on different levels of induced rarity.
To do this we gradually rarified the minority behaviors in each dataset artificially, by subsampling the abnormal class, before starting a pass of the pipeline. 
For each rarity level, we repeat the full pipeline, including all labeling efforts.
Since these experiments are highly stochastic we repeat them with different random seeds, using 9 or 3 replicates for the synthetic or biological data, respectively.

\begin{figure}[htbp]
  \centering
    \includegraphics[scale=0.4]{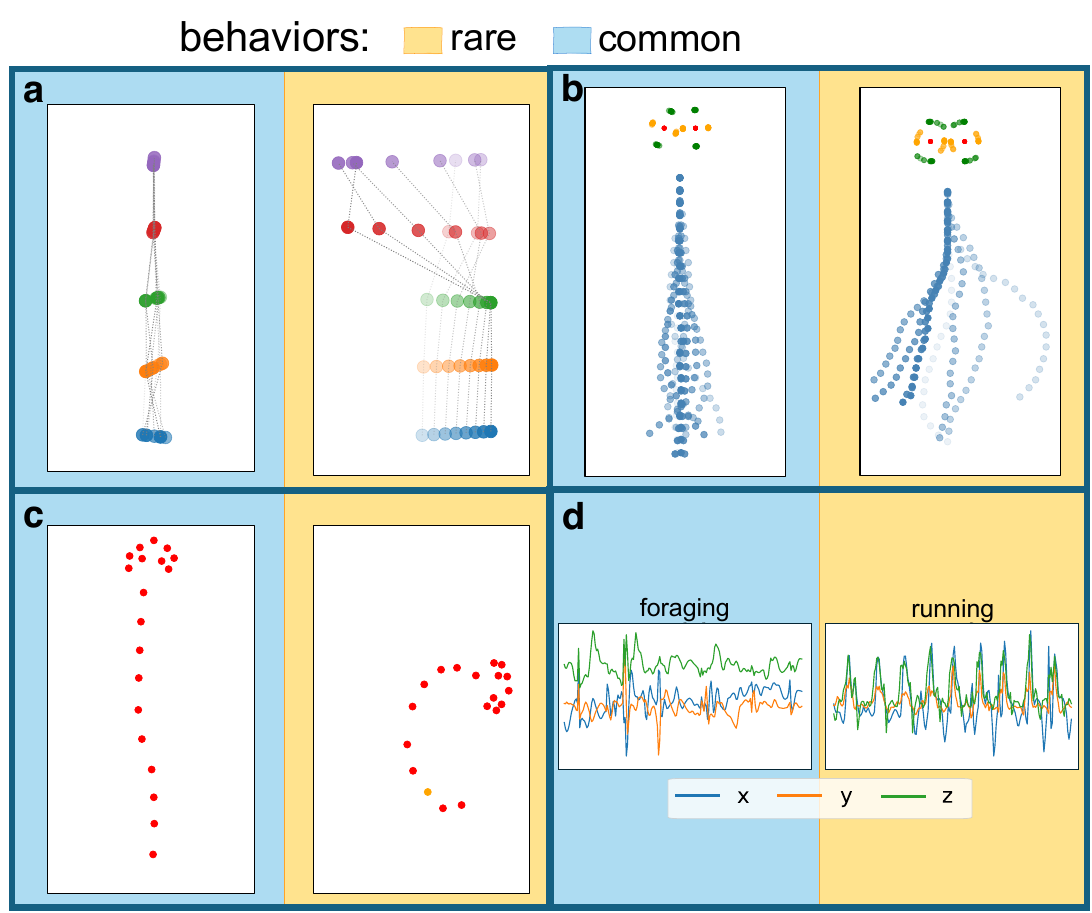}
    \caption{Data examples. The common behaviors are on the left (blue), and the rare behaviors are on the right (yellow) for each panel.  \textbf{(a)} Samples from the synthetic dataset ({\scriptsize$\mathit{behavior SD} = 0.5$}), depicted as an overlay of all frames in the sequence (lighter color for earlier frames). \textbf{(b)} Two (of 5) behaviors from FishLarvae1: exploration (common), and strike (rare), depicted as an overlay of all frames. \textbf{(c)} Two (of 8) behaviors from PoseR: scoot (common), and slc (rare), a single representative frame. And \textbf{(d)} Two (of 4) Accelerometry samples from Meerkat: foraging (common), and running (rare), depicted as acceleration (y-axis) against time (x-axis) for different directions (color). 
    The data are highly variable in terms of motion behaviors, presenting the real-world challenges of discerning rare behaviors in research data.
    }
    \label{figure:data_samples}
\end{figure}

\subsection{Datasets}
\paragraph{Data representation.}
We use a graph-based representation for all our datasets. Aside from one dataset, all of our datasets are pose estimation datasets, where each vertice corresponds to a landmark of interest on an animal (a real or synthetic fish) with two channels - x and y.
The relationships between the vertices are defined by an adjacency matrix, which defines the "skeleton" of the animal.
One dataset, the Meerkat dataset, is an accelerometry dataset (see below). The raw data in this dataset is accelerations on each of the axes (see \Cref{figure:data_samples}d). 
To create a representation similar to a pose graph from these one-dimensional accelerations we use the accelerations on the (xy, yz, xz) planes to create a graph with three two-channeled vertices.

\begin{figure}
\centering
\includegraphics[scale=0.2]{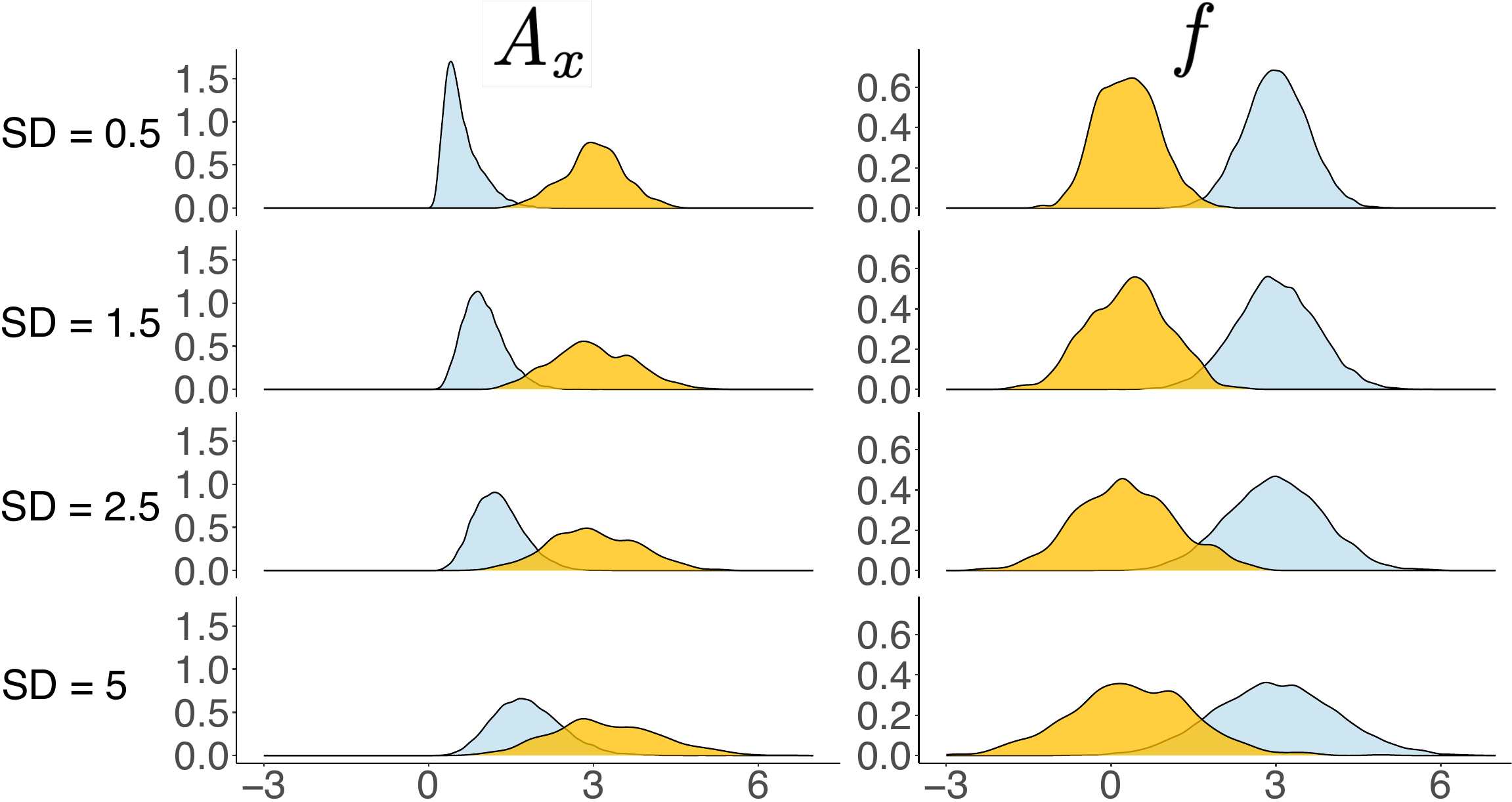}
\caption{Movement parameters for the synthetic dataset. Density plots of the two parameters used to generate keypoint movement. Each keypoint's movement was modeled as a sinus with amplitude A (left columns) and frequency f (right column). For the common class (blue) $\mathit{f > A_x}$, and the opposite for the rare class (yellow). We controlled the similarity between the rare and common "behaviors" by changing the standard deviation of each of the multivariate Gaussians 
(SD, rows), similarity grows with this behavior SD.}\label{fig:syntheticparams}
\end{figure}

\subsubsection{Synthetic data}
To test our ability to detect rare behaviors in a controlled setting, we created a synthetic dataset with two "behaviors" that differ between them by a simple kinematic rule. We draw inspiration from the swimming behavior of real fish (see \Cref{figure:data_samples}).
Each sample is comprised of a 5-keypoint pose sequence of 9 timesteps. 
The movement of each keypoint is determined by a sinus with some amplitude $\mathit{A}$ and frequency $\mathit{f}$, each is drawn from a separate multivariate Gaussian distribution with means $\mathit{\hat{A}}$ and 
$\mathit{\hat{f}}$ respectively. 
The behaviors differ in that for the common behavior {\small$\mathit{\hat{f}} > \mathit{\hat{A}}$} and vice versa for the rare behavior (\Cref{fig:syntheticparams}).
For the test set, we changed the means of $\mathit{\hat{A}}$ and $\mathit{\hat{f}}$ to test whether the model learned the kinematic rule.

Using this simple scheme, we create datasets with different levels of similarity between the rare and common behavior (\Cref{fig:syntheticparams}).
We change the similarity between behaviors by increasing the standard deviation of the multivariate Gaussian, allowing the behavior to be more variable. 
We think this is a biologically relevant way to simulate similarity between behaviors, as behaviors in nature are highly variable.
We set 4 levels of the standard deviation, where the lower the value the more distinct the behaviors ({\scriptsize $\mathit{behavior SD} = [0.5, 1.5, 2.5, 5]$}, \Cref{fig:syntheticparams}).
We set the frequency of the rare behavior in the dataset to be 5\% and test the effect of different frequency levels on performance in an ablation study (Section 5.2).

\subsubsection{Biological datasets}
We used three published datasets documenting freely behaving animals sourced from experimental neurobiology and field ecology \cite{chakravarty2019novel,johnson2020probabilistic, mullen2023poser}. 
As in most natural datasets, the behavior classes follow a long-tail distribution where most samples belong to a few common behaviors.
In this study, we seek to find the rare behavior classes at the long tail of the distribution. 
Hence, we divide each dataset into "normal" for the common classes 
and "abnormal" or rare otherwise.
Thus, our common and rare classes encompass inter- and intra-behavior variability.

We briefly introduce the datasets below and give the unnamed datasets (\cite{johnson2020probabilistic, chakravarty2019novel}) names we'll use for the rest of the paper. We visualize sample data in \Cref{figure:data_samples}.

\begin{enumerate}

\item \textbf{FishLarvae1} - Johnson et al \cite{johnson2020probabilistic} created a dataset of pose-estimated sequences of larval Zebrafish (\textit{Danio rerio}) chasing prey in a large laboratory arena. 
Behaviors were divided into 5 main categories - explore, pursuit, j-turn, abort, and strike. 
We took the former two to be normal (making up $88\%$ of the dataset) and the latter three to be abnormal ($12\%$ of the dataset).

\item \textbf{PoseR} - Mullen \etal \cite{mullen2023poser} present another pose estimated larval Zebrafish behavioral dataset. 
The data is compiled from several separate neurobiological experimental assays. Behaviors were divided into - burst swim, routine turn, j-turn, scoot, long-latency C-bends, slow-latency C-bends, O-bends, and noise. 
We considered the first four behaviors normal and the latter four as rare ($95.7\%$ and $4.3\%$ of the dataset, respectively).

\item \textbf{Meerkat} - We used an accelerometry dataset of meerkat behavior \cite{chakravarty2019novel} to show that our method generalizes to other data modalities and species. 
The dataset was assembled by fitting the animals with tri-axial accelerometers while simultaneously monitoring them with video in the wild.
The dataset contains four behaviors - foraging, vigilance, resting, and running. We considered the former two as normal and the latter two as rare ($94.4\%$ and $5.6\%$ of the dataset, respectively).
\end{enumerate}

See supplementary Section S1.2 for details on the training splits used.
In all three studies, the original problem tackled was multi-class behavior classification; thus, our results are not directly comparable with their benchmarks.

\subsection{Performance metrics and statistical analysis}

To assess our labeling scheme, the performance of each sampling method (ours or random) was measured as the performance of a classifier trained using each of the resulting datasets. 
We use the Area under the Precision-Recall curve (AuPRC, range 0-1) to evaluate the classifier performance, a common metric for unbalanced data \cite{saito2015precision}.
The AuPRC expected under a random classifier equals the fraction of the positive class in the data. Thus, this metric is sensitive to data imbalance and cannot be used to directly compare performance between test sets with different data imbalances \cite{saito2015precision, brabec2020model}.
For this reason, we kept the same test set for each dataset through all rarity experiments. 

To statistically analyze and visualize our results, we modeled AuPRC as a function of rarity level (log-transformed), labeling effort, and sampling method. For the synthetic dataset, we added behavior SD. 
To assess our method we plot the AuPRCs predicted by the model as a function of rarity for each labeling method (Section S2).

\subsection{Implementation Details}

For STG-NF, we used the code implementation provided by \cite{hirschorn2023normalizing} in the PyTorch \cite{pytorch2019} framework. We used {\scriptsize $\mathit{K}=4$, $\mathit{L}=3$, $\mathit{R}=3$}, 16 hidden channels, and a batch size of 256. 
All other hyperparameters slightly varied between datasets.
The STG-NF models are lightweight with $~8-10.5K$ parameters, depending on the dataset configuration.
They were trained for 4 epochs as we found it was enough for models to converge.
For ST-GCN we used an implementation provided in the STG-NF codebase. In the rarity experiments, we trained the model for 2 epochs, using a batch size of 256. 

Any modifications, all hyperparameters, and all experiments we ran are available in our codebase: \url{https://github.com/shir3bar/SiftingTheHaystack}.
The synthetic dataset we created is publicly available at: \url{https://doi.org/10.5281/zenodo.14266407}.
All statistical analyses were conducted in R, and all performance metrics were calculated using the R precrec package \cite{saito2017precrec} and emmeans \cite{russel2024emmeans}. The raw data used for these analyses are publicly available at: \url{https://doi.org/10.5281/zenodo.14253658}.

\begin{figure*}[htpb]
  \centering
    \includegraphics[scale=0.6]{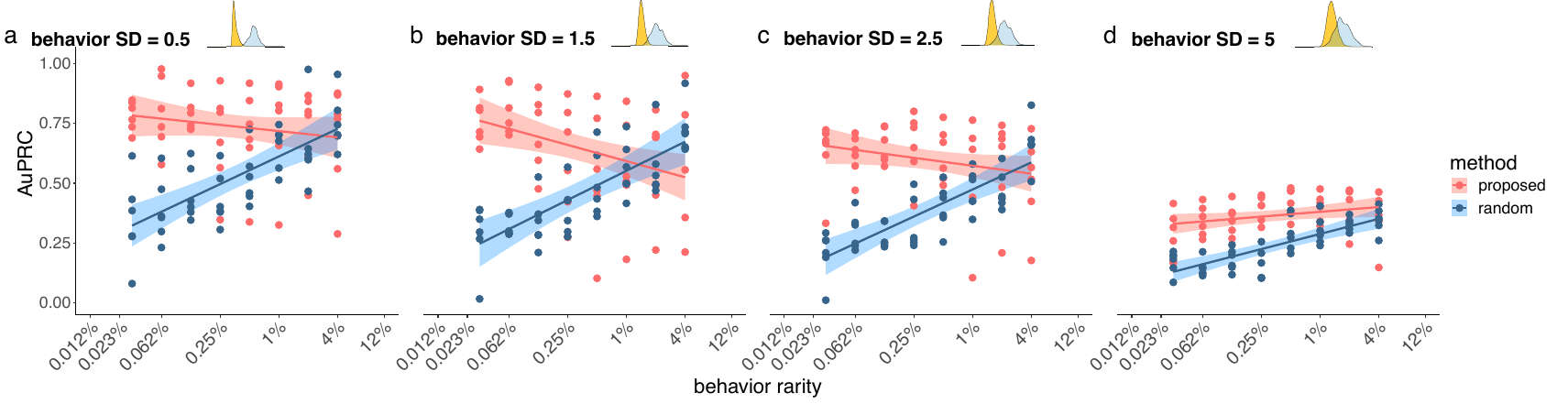}
    \caption{Labeling pipeline performance on the \textbf{synthetic data}. Predicted performance (y-axis) of a classifier under different induced rarities (x-axis) for a labeling budget of {\scriptsize$\mathit{N_{reviewed}}=200$}. We compare our anomaly detection-based labeling approach (red), to randomly picking and labeling samples from the dataset (blue). Panels (a-d) are organized in order of increasing behavioral similarity. Our labeling method provides stable and better performance across a range of artificial rarities and at all levels of behavioral similarities. However, the method yields the largest benefit at higher rarities (left side of the x-axis).}
    \label{figure:simulated_visreg}
\end{figure*}

\section{Results}
We start by motivating our approach using results from the synthetic data, which can be viewed as a controlled experiment where we probe both model and pipeline performance.
We then present the application of our method to 3 biological research datasets.

\begin{table}[htbp]
\centering
\resizebox{\columnwidth}{!}{
\begin{tabular}{llll}
\toprule
&& \multicolumn{2}{c}{labeling method} \\
&&\multicolumn{2}{c}{\small($\mathit{mean AuPRC \pm sd}$)} \\ 
\cmidrule(lr){3-4}
data type & dataset & proposed & random \\ 
\midrule\addlinespace[2.5pt]
\multirow{4}*{synthetic} & $\mathit{behaviorSD} = 0.5$ & \textbf{$0.74\pm0.049$ }& $0.51\pm0.22$ \\ 
& $\mathit{behaviorSD} = 1.5$  & \textbf{$0.65\pm0.13$}& $0.45\pm0.23$ \\ 
& $\mathit{behaviorSD} = 2.5$  & \textbf{$0.6\pm0.063$} & $0.37\pm0.21$ \\ 
&$\mathit{behaviorSD} = 5$  & \textbf{$0.36\pm0.038$} & $0.23\pm0.12$ \\ 
\hline\addlinespace[2.5pt]
\multirow{3}*{biological}  & FishLarvae1 & \textbf{$0.27\pm0.067$} & $0.21\pm0.022$ \\ 
& PoseR & \textbf{$0.77\pm0.11$} & $0.49\pm0.21$ \\ 
& Meerkat & \textbf{$0.47\pm0.22$} & $0.25\pm0.15$ \\ 
 
\bottomrule
\end{tabular}}
\caption{Performance of labeling pipeline.
The estimated performance of the labeling pipeline in AuPRC across all rarity levels for a labeling budget of {\scriptsize $\mathit{N_{reviewed}} = 200$}. 
Our method is more beneficial than random sampling across all datasets.}\label{table:all_auprcs}
\end{table}

\vspace{-10pt}

\subsection{Synthetic data}

For a given labeling budget, our proposed labeling method performed better than random sampling, particularly when rarity levels were higher (\Cref{figure:simulated_visreg}).
The mean improvement on random sampling was $64.82\%\pm 2.36$ {\small($\mathit{mean} \pm \mathit{SE}$)}, 
across all rarity levels and labeling efforts. 
The advantage given by our method was variable, changing across behavior similarity (low to high, \Cref{figure:simulated_visreg}a-d), labeling efforts, and rarity level (along the x-axis of \Cref{figure:simulated_visreg}).

\textbf{Effect of labeling budget.} The labeling effort had a small (but statistically significant) effect on performance for both methods (Figure S1). In the main text, we discuss the results for a labeling budget of {\small$\mathit{N_{reviewed}} = 200$}, but note that changing the labeling effort does not change the overall trend.

\textbf{Effect of behavior similarity.} Increased behavior similarity (i.e., high SD) hurt the performance of both methods, however, our method still dominated. 
Mean AuPRCs dropped from $0.79$  AuPRC to $0.40$ in our method and for the random sampling from $0.59$ to $0.28$, from most distinct to most similar behaviors respectively (\Cref{figure:simulated_visreg}, \Cref{table:all_auprcs}) 

\textbf{Effect of rarity level.} The effect of rarity level was stronger in the random method ({\scriptsize$\mathit{slope} = 0.17\pm0.05$, $\mathit{mean}\pm \mathit{sd}$, for all behavior SDs}). 
That means that for every order of magnitude increase in behavior rarity (\eg, from 1\% to 0.1\% of the dataset) the random method loses 0.17 points from the AuPRC score. 
Our method was mostly unaffected by the rarity of the synthetic behavior ({\scriptsize$\mathit{slope}=-0.04\pm0.06$, $\mathit{mean}\pm \mathit{sd}$}).
When behavior rarity is low (i.e., has a higher \% in the dataset, right side of the x-axis) we see that random sampling becomes as effective as our method. 
We believe the reason for this lies in the increased chance of finding rare instances that lie closer to the decision boundary of the classifier. 
We discuss this in the Limitations section below.

Often, researcher have no exact apriori knowledge of the degree of similarity or level of rarity within their datasets. 
This investigation serves to test the boundaries of our method. Though we find the performance is still far from a fully supervised benchmark (Section 5.2), we deliver good performance on a limited labeling budget. 

\subsection{Application to biological datasets}
\begin{figure*}
  \centering
            \includegraphics[scale=0.6]{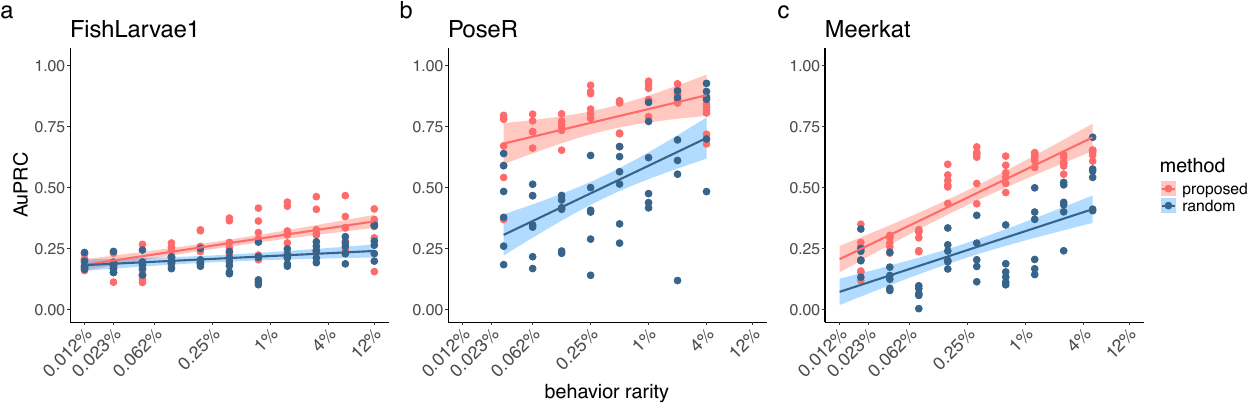}
        \caption{Labeling pipeline performance on \textbf{biological data}. Predicted performance (y-axis) of a classifier under different induced rarities (x-axis) for a labeling budget of {\scriptsize$N_{reviewed}=200$}, 
        for the biological datasets (panels a-c). We compare our anomaly detection-based labeling approach (red), to randomly picking and labeling samples from the dataset (blue). Trends differ between datasets, however, our method provides a clear advantage in two datasets (b,c) across most of the rarity range.}
        \label{figure:real_visreg}
           

\end{figure*}

The performance of the labeling pipeline improved on random sampling also in the biological datasets (\Cref{table:all_auprcs}). 
Though our method was always superior, the advantage was variable across datasets (\Cref{figure:real_visreg} a-c), labeling effort, and rarity level (\Cref{figure:real_visreg}, x-axis). 

\textbf{Effect of labeling effort.} As in the synthetic datasets, labeling effort had a small but significant effect on performance which varied between datasets (Figure S2).
For the FishLarvae1 dataset, the labeling effort had the strongest effect on performance, in that a researcher would need to review at least 300 clips to get a decent classifier ({\scriptsize$\mathit{AuPRC}=\sim0.45$}). For the PoseR and Meerkat datasets performance stabilizes after only 100 reviewed clips.
In the main text, we discuss results for a labeling effort of {\small$\mathit{N_{reviewed}} = 200$}.

\textbf{Effect of rarity level.} The performance as a function of rarity had two different trends. 
For the PoseR and Meerkat datasets, our method improved performance substantially (\Cref{table:all_auprcs}, rows 3 and 4, respectively).
The performance improvement increased with rarity, similar to the synthetic datasets (\Cref{figure:real_visreg}b, and c, respectively). 
In PoseR, like in the synthetic data, this trend was driven by the high dependency of random sampling on rarity (\Cref{figure:real_visreg}b). 
In Meerkat, however, both methods were similarly affected by rarity (\Cref{figure:real_visreg}c).
On the other hand, for the FishLarvae1 dataset performance improved when rarity decreased ({\scriptsize$\mathit{proposed\; method\; slope} = 0.08$}, \Cref{figure:real_visreg}a), but overall improvement was lowest (\Cref{table:all_auprcs}, row 2). 

Though some patterns were similar to those observed in the synthetic data, 
biological datasets are, predictably, much more challenging. This performance gap may be attributed to mislabeled samples, noisy samples, or higher variance typical of natural behavior. Our method truly shines when the rare behavior frequency is {\small$\leq 1\%$} of the dataset, enabling the user to train rare behavior classifiers that yield a substantial advantage compared to traditional random sampling.

\subsection{Ablation studies}
\textbf{What would performance be if we labeled the entire dataset?} We benchmark the classifier and the anomaly detector on the complete synthetic and biological datasets (Figures S3 and S4).
For the synthetic dataset, both models' performance was adversely affected by behavior similarity. 
This effect was more pronounced in the anomaly detector, with performance dropping to 0.19 mean AuPRC at the highest similarity level compared to 0.5 for the classifier (Figure S3).
For the biological datasets, we find that ST-GCNs are highly performant ({\scriptsize$\mathit{mean\; AuPRC} = 0.97$}) when given a large labeled dataset, even in the face of high data imbalance. 
The anomaly detector showed decent results only on one of the datasets ({\scriptsize$\mathit{mean\; AuPRC} = 0.43$}, Figure S4).

\textbf{Effect of the baseline rarity level on models.}
We modulated the baseline frequency of the rare behavior in the synthetic dataset to see if this affects model performance when trained on the full dataset.
We tested 4 rarity levels, where the frequency of the rare behavior is changed ({\small $\mathit{\%rarity} = [1.5\%, 5\%, 12\%, 24\%]$ of the data}). 
Table S2 summarises the results across all rarity levels and behavior similarities. 
The rare behavior frequency by itself did not affect performance. However, rarity had a combined effect with behavior similarity, such that when behaviors were more fine-grained, rarity had a stronger negative effect on performance (see statistical analysis in section S2).

\section{Limitations}\label{sec:limits}
As seen in \Cref{figure:simulated_visreg,figure:real_visreg} our method is most beneficial when the rare behaviors make up {\small$\leq 1\%$} of the dataset. 
Above it, random sampling finds samples easily enough and has a greater chance of catching hard positives (i.e., rare behaviors with low anomaly scores).
In our pipeline, we can identify hard negatives, since we sample a wider region of the anomaly score distribution tail (\Cref{figure:overview}). However, since we do not review samples drawn from the center of the distribution, we exclude the possibility of 
finding hard positives. 
Mullapudi \etal \cite{mullapudi2021learning} in their active learning scheme observed that, at some point, the benefits of sampling easy positives diminish. They proposed an adaptive approach where the samples to be reviewed in each iteration shift between positive and negative regions according to their relative class sizes.
This adaptive sampling, as well as an active learning setting, is an interesting future direction for our method.

Graphs proved to be beneficial for studying behavior kinematics. However, since they abstract away all color and appearance information in the recordings, they are not suitable for studying behaviors associated with color/appearance.

The unsupervised model struggled with the FishLarvae1 dataset \cite{johnson2020probabilistic} (Figure S4a).  An investigation of model performance led us to conclude that for many instances, the difference between the rare and common behaviors lies in the movement of the 8 eye coordinates (see \Cref{figure:data_samples}b), while the rest of the 23 body vertices were similar. We observed the anomaly detector failed to learn this fine-grained difference, which reduced the 
performance of the entire pipeline.

We did not test generalization to new individuals in our pipeline. Rather, our method assumes you have all of the recording data available and would like to mine within it for rare behaviors. Due to the scarcity of data, we worked without a validation set (Section S1.2).

\section{Conclusions}
In this study, we tackled a common challenge for biologists, that of finding rare behavior instances in a large collection of recordings. We show that by using anomaly detection one can quickly annotate a rare behavior dataset which can then be used to train a rare behavior classifier.
Our approach does not make any assumption about the type of rare behaviors to be found, nor about how many different types there are or about the number of rare instances. 
Through this work, we also find that graphs are an excellent abstraction for studying the motion patterns of behaviors, providing a strong prior for the model and yielding good classification results using a small number of samples.

\paragraph{Acknowledgments:} Parts of this research were supported by the Israeli Science Foundation (ISF) grant 2123/23 to SA, and ISF grant 592/22 to RH. SB thanks the RH lab for being awesome.

{\small
\bibliographystyle{ieee_fullname}
\bibliography{references}
}

\end{document}


\tableofcontents
\section{Data details} 
\subsection{Synthetic data generation} 
Each synthetic sample is comprised of a 5-keypoint pose sequence of 9 timesteps. 
The movement of each keypoint is determined by a sinus with some amplitude A and frequency f, each is drawn from a separate multivariate Gaussian distribution with mean $\hat{A}$ and mean $\hat{f}$ respectively. 
The behaviors differ in that for the common behavior $\hat{f} > \hat{A}$ and for the rare behavior $\hat{A} > \hat{f}$ (see main text Figure 4).
For the Y coordinate, to keep the motion relatively simple, we set $\hat{A}$ to be a small perturbation, varying the movement mostly in the X coordinate of each vertice. 
This qualitatively made samples appear more similar to the movement in the biological fish datasets we used (see main text Figure 3).

The amplitudes and frequencies of keypoints on the same sample are weakly correlated (covariance = 0.3).
For the train set we set $\hat{A}_1 = 0.6$ and $\hat{f}_1 = 6$ for the common behavior and $\hat{A}_2 = 6$ and $\hat{f}_2 = 0.6$ for the rare behavior.
For the test set we set $\hat{A}_1 = 0.3$ and $\hat{f}_1 = 3$ for the common behavior and $\hat{A}_2 = 3$ and $\hat{f}_2 = 0.3$ for the rare behavior.
We changed the means of the parameters to test whether the model learned the kinematic rule.

We created a separate dataset for each behavior similarity level. 
We also tested the effect of the baseline rarity level in the dataset on pipeline performance. To do that, we generated datasets with different initial rarity levels - ($\%rarity = 1.5\%,5\%,12\%,24\%$ of the dataset).
We created a total of 16 datasets (4 behavior similarities X 4 rarity levels). The number of samples in each dataset varied slightly due to the rarity modulation. All dataset sizes are summarised in Table S1.
The code to generate the data is provided in our \href{https://github.com/shir3bar/SiftingTheHaystack}{codebase}, and the exact datasets used are provided in the following data repository: \href{https://doi.org/10.5281/zenodo.14266407}{10.5281/zenodo.14266407}.

\begin{table}[htbp]
\centering
\resizebox{\textwidth}{!}{
\begin{tabular}{llrrrrrrrr}
\toprule
 &  & \multicolumn{3}{c}{train} & \multicolumn{3}{c}{test} \\ 
\cmidrule(lr){3-5} \cmidrule(lr){6-8}
data\_type & name & normal & abnormal & total & normal & abnormal & total \\ 
\midrule\addlinespace[2.5pt]
synthetic & 5\% rarity & 30000 & 1500 & 31500 & 10000 & 525 & 10525 \\ 
 & 1.5\% rarity & 30000 & 450 & 30450 & 10342 & 158 & 10500 \\ 
 & 12\% rarity & 30000 & 3600 & 33600 & 9240 & 1260 & 10500 \\ 
 & 24\% rarity & 30000 & 7200 & 37200 & 7980 & 2520 & 10500 \\ 
biological & FishLarvae1 & 111985 & 15412 & 127397 & 22546 & 3545 & 26091 \\ 
 & PoseR & 25780 & 1156 & 26936 & 4557 & 202 & 4759 \\ 
 & Meerkat & 58477 & 3435 & 61912 & 11667 & 716 & 12383 \\ 
\bottomrule
\bottomrule
\end{tabular}}
\caption{Number of samples for each dataset.
Number of normal, abnormal, and total samples for the train and test sets of each dataset. For the synthetic datasets, sample sizes were the same for different similarity levels within the same rarity level.}\label{table:all_ns}
\end{table}

\subsection{Biological datasets - details on datasets and splits} 

\subsubsection{FishLarvae1}
Johnson et al. \cite{johnson2020probabilistic} documented larval Zebrafish (\textit{Danio rerio}) chasing prey in a large arena in the laboratory. Videos were shot by tracking the fish across an arena with an overhead camera at 60 frames per second (FPS). 
Videos were subsequently pose-estimated and then segmented into clips according to the behaviors and thus have varying durations. The pose sequence clips were made publicly available.
Behavioral labels were assigned using unsupervised behavioral clustering and divided by researchers into 5 main behavioral categories - explore, pursuit, j-turn, abort, and strike. 

The behavior clips were segmented by the annotator such that the first frame is the start of the behavior and the last frame is the end. Such that the behavior instances had varying numbers of timesteps (or frames). This is unlike the two other datasets which have a set duration for all samples.
The original dataset included several frames preceding or succeeding the behavior, but we ignored these as they had no behavior annotations.

We note that larval fish striking behavior is very fast ($\sim40$ milliseconds), and the framerate this dataset was acquired in is thus sub-optimal to document the behavior. For comparison, the PoseR dataset (below) was acquired at 300 FPS. Indeed, when we reviewed examples from the dataset, the characteristic s-shaped posture preceding a strike appeared only in a single frame. This makes the distinction between behaviors particularly challenging.

\paragraph{Data splitting} The data had been acquired in several long filming sessions (trials) and repeatedly from roughly 100 different individual fish. Individual trials may be on different days. 
Each behavior sequence, i.e., an annotated pose sequence, has an associated trial ID and individual ID. 
We split the data into test and train such that clips from the same filming trial are all in the same partition (either test or train) however we didn't consider individual identity as each individual had multiple trials. 
We provide code for data preparation in our \href{https://github.com/shir3bar/SiftingTheHaystack/tree/main/data_prep}{codebase} which includes the data splits we used.

\subsubsection{PoseR dataset}
Mullen et al. \cite{mullen2023poser} present another larval Zebrafish behavioral dataset. 
The data is compiled from several separate neurobiological experimental assays. 
Unlike the previous dataset, here the motion of the larvae is restricted to a small $25mm\;x\;25mm\;x\;25mm$ aquarium.  
Videos were acquired at 300 FPS and then pose-estimated; these pose sequences were made publicly available. 
Each behavior sequence was 1 second, i.e., 300 frames.
Behavioral labels were assigned using unsupervised behavioral clustering and divided into - burst swim, routine turn, j-turn, scoot, long-latency C-bends, slow-latency C-bends, O-bends, and noise. 

\paragraph{Data splitting and cleaning}
Even though the dataset had a dedicated "noise" category, we found that in many cases skeleton vertices would flicker and be estimated far from the rest of the skeleton.
At the same time, we found that our anomaly detector is particularly good at finding these samples and assigning them a high anomaly score. 
While potentially useful, this was not what we attempted to do in this study.
So we filtered the PoseR dataset by dropping frames that had landmarks that were above a threshold distance from the rest of the fish. 
Since the dataset was acquired at a high frame rate, dropping a single frame did not affect the smoothness of the movement. 
However, we took a conservative approach, and if a clip had more than 50 frames non-consecutive dropped, or more than 20 consecutive frames dropped we completely removed the clip from the data. 
In total, we removed 1976 clips from the train data and 344 from the test data using this method.

As for data splitting into train and test, we used the same splits used in the original paper \cite{mullen2023poser}. 
The cleaning code with the data preparation code is available in our code repository. 

\subsubsection{Meerkat}
We used an accelerometry dataset acquired by Chakravarty et al. \cite{chakravarty2019novel} to show the generality of the framework to other data modalities and species. This is a dataset of Meerkat behavior assembled by fitting the animals with collars mounted with tri-axial accelerometers while simultaneously monitoring them with video. Acceleration data was acquired at 100 Hz/axis and split into two-second clips resulting in 200 "frames" with 3 data points per frame for each sequence. 
The behavioral labels were determined from the videos and divided into four behaviors - foraging, vigilance, resting, and running.

\paragraph{Data splitting and preparation}
As described in the main text, to make the data compatible with the anomaly detection framework we used we had to make each vertice have two channels. We thus took the planar accelerations along (xy, xz, yz). Data preparation code is provided in our codebase.
Each behavior segment was associated with the ID of the filmed individual.
A total of 10 meerkats were filmed in 11 filming sessions.
Though we initially tried to separate individuals, the task proved challenging as it significantly changed the distribution of behaviors in the dataset.
Thus we split the dataset randomly into test and train. Given that our goal in the end is to find rare behaviors within a single dataset efficiently, we feel this does not hurt our evaluation. 
The splits we used are provided in the data preparation code.

\section{Statistical analysis methods and results} 
\subsection{Statistical modeling of rarity experiments}
To understand the effects of induced rarity level and labeling effort on our method, and robustly compare it to random sampling, we modeled the rarity experiments for each dataset using linear models in R. 
The code statistical analysis is available in our codebase. The files containing the raw outputs of the experiments are provided in the following data repository: \href{https://doi.org/10.5281/zenodo.14253658}{10.5281/zenodo.14253658}

Below we provide a short paragraph describing the main findings of this analysis.
When reading the outputs of such models we look at the coefficients each variable is assigned to assess its effect on performance, and at the p-value (p) to assess whether this effect is statistically significant.

\paragraph{Synthetic data}
We modeled the AuPRC as a function of sampling method (ours or random), rarity level (log-transformed), and labeling effort, including a three-way interaction between these parameters. This interaction term essentially means that different trends may appear in the data with different combinations of these parameters.

We considered each of the 4 behavior SDs separately and created a separate model for each.
The results show that both methods are affected by behavior rarity but in different ways. 
While our method has a weak negative correlation with the frequency of the rare behavior (i.e., performance increases when behavior is rarer, coefficients between -0.18 - 0.03), random sampling performance is positively correlated and with a stronger effect (coefficients between 0.25-0.119).
The results show that our method provides more stable performance across the range of rarities and similarity levels.

In Figures 4 and 5 in the main text, we use the statistical models of each dataset to plot the estimated performance given a set labeling budget of 200 samples at different rarity levels. This is done using the 'visreg' package in R \cite{visreg}. For Table 1 in the main text, we similarly use the model of each dataset to calculate the estimated mean performance across all rarity levels using the 'emmeans' package in R \cite{emmeans}.

\begin{figure*}[htbp]
  \centering
    \includegraphics[scale=0.28]{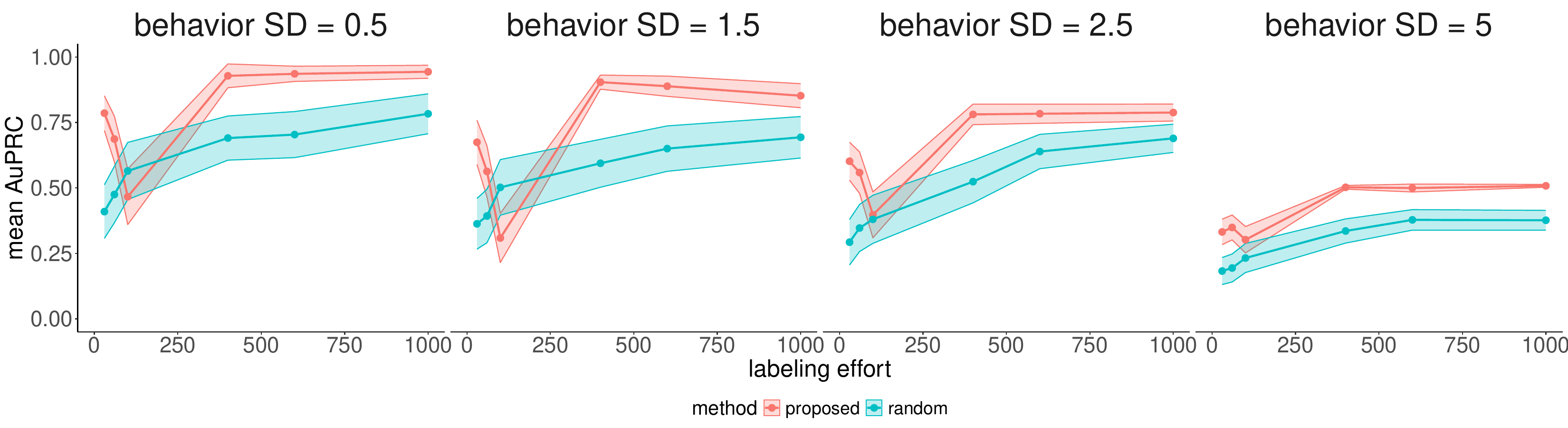}
    \caption{Effect of labeling effort on performance for the \textbf{synthetic} datasets. 
    Performance (y-axis) as a function of labeling effort (x-axis) at different behavior similarities (behavior SD, a-d) for our method (red) and the traditional method (blue). Performance was measured in AuPRC and averaged across all tested rarities, the ribbon represents the upper and lower confidence intervals (95\% CI). Our method was superior for all labeling efforts, and saturated at around 300 reviewed clips.} 
    \label{figure:label_effort_sim}
\end{figure*}
\begin{figure*}[htbp]
  \centering
    \includegraphics[scale=0.27]{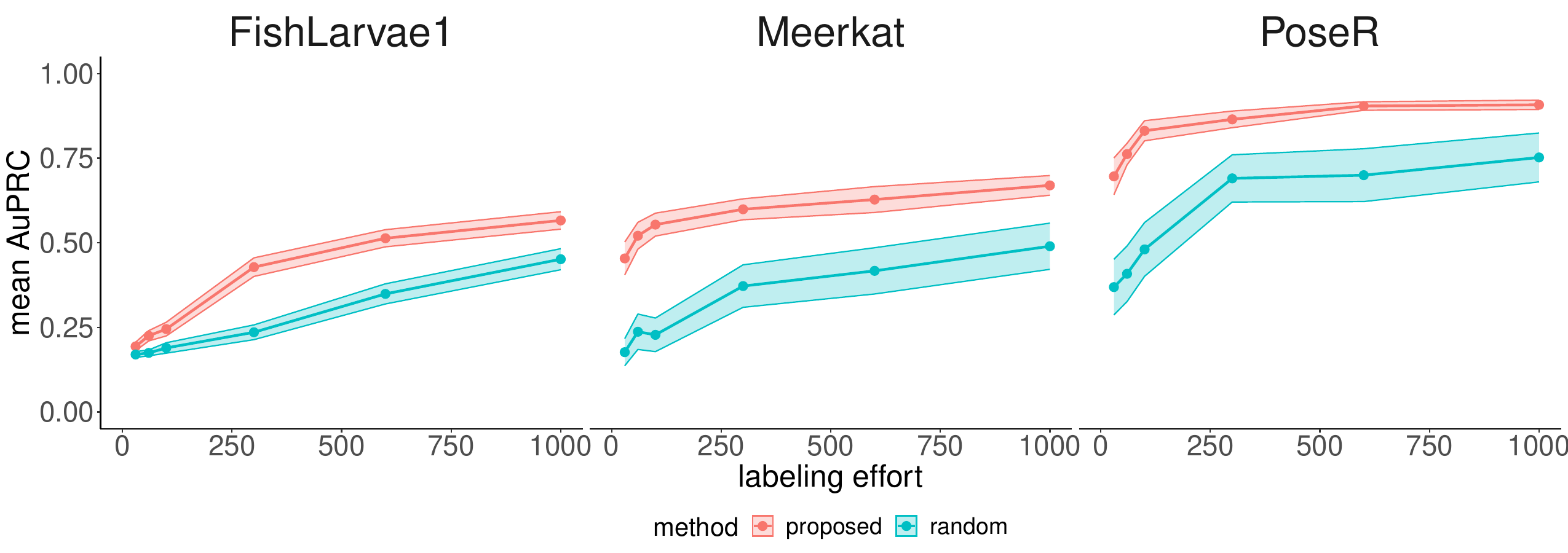}
    \caption{Effect of labeling effort on performance for the \textbf{biological} datasets. 
    Performance (y-axis) as a function of labeling effort (x-axis) for the different datasets (a-c) for our method (red) and the traditional method (blue). Performance was measured in AuPRC and averaged across all tested rarities, the ribbon represents the upper and lower confidence intervals (95\% CI). Our method was superior for all labeling efforts and saturated after 300 reviewed clips for the FishLarvae1 dataset and only 100 reviewed clips for the PoseR and Meerkat datasets.} 
    \label{figure:label_effort_real}
\end{figure*}

\subsubsection{Effect of labeling effort}
The labeling effort had a small but significant effect on performance for both methods (labeling effort between 30-1000, coefficients for both methods between -0.0004 - 0.0012). 
In Figures \Cref{figure:label_effort_sim} and \Cref{figure:label_effort_real} we plot the actual (i.e., not estimated) mean AuPRCs as a function of labeling effort for the synthetic and biological datasets, respectively.
Both figures show similar trends, our method not only yields better performance overall but also does so using a smaller labeling effort. This can be seen by looking at when the curve starts to plateau.

\subsection{Statistical analysis of ablation studies}
\subsubsection{Synthetic data}
To statistically evaluate the effect of the behavioral similarity and the frequency of the rare behavior on model performance given the fully labeled dataset (i.e., model benchmarking) we used a linear model. 
We modeled standardized Area under the Precision-Recall Curve (AuPRC) as a function of behavioral similarity and frequency and included their interaction and the architecture.

\paragraph{Standardized AuPRC}
The AuPRC expected under a random classifier is equal to the fraction of the positive class in the data. Thus, this metric is sensitive to data imbalance and cannot be used to directly compare performance between test sets with different data imbalances \cite{saito2015precision, brabec2020model}.
To standardize the AuPRC we calculate the difference between the AuPRC and the expected performance (which is equal to the \%rarity). 
Because this procedure alone will bias the metric against high behavioral frequencies (with higher expected performance) we further divide the quantity by the maximum possible difference between AuPRC and expected performance \ref{eq:stdauprc}. 
This yields a metric that, like AuPRC, ranges from 0 for poor performance to 1 for perfect performance.
\begin{equation}
    standardised\; AuPRC = \frac{AuPRC - \%\:rarity}{1- \%\:rarity}
\end{equation}\label{eq:stdauprc}

\begin{figure*}[htbp]
  \centering
    \includegraphics[scale=0.3]{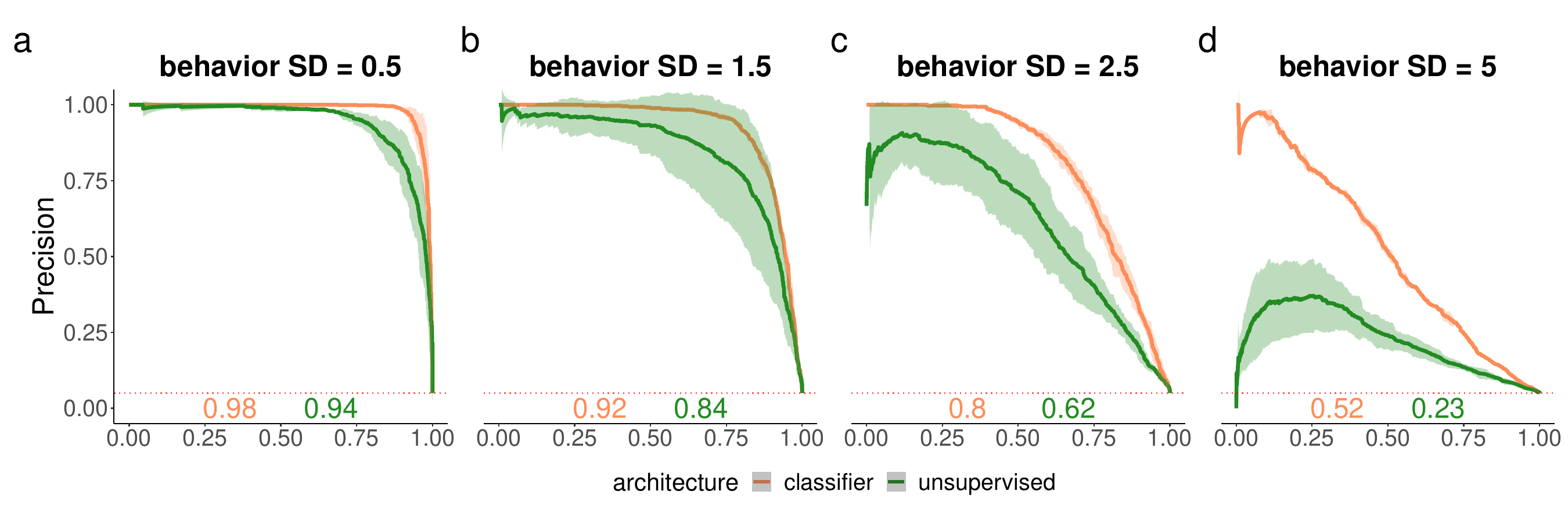}
    \caption{Example of architecture performance on \textbf{synthetic data}. Precision-Recall curves for the 5 \% behavior rarity dataset for different behavioral overlaps (panels a-d). Colored numbers in each pane correspond to the respective area under each of the curves. ST-GCN classifiers (orange) dominate performance.
    Unsupervised Normalizing Flows (green) are competitive for lowest behavioral similarities but their performance degrades more quickly.}
    \label{figure:sim_prcs}
\end{figure*}

\begin{table}[htbp]
\centering
\resizebox{0.8\textwidth}{!}{
\begin{tabular}{|l|l|llll|}
\toprule
\multicolumn{2}{|l}{} & \multicolumn{4}{|c|}{Behavior Standard Deviation (more overlap $\rightarrow$)} \\ 
\cmidrule(lr){3-6}
\multicolumn{1}{|l|}{Method} & \multicolumn{1}{l|}{\% Rare} & 0.5 & 1.5 & 2.5 & 5 \\ 
\midrule
\multirow{4}{*}{Anomaly detector (STG-NF)} & 1.5\% & \cellcolor[HTML]{9AD93D}{\textcolor[HTML]{000000}{0.85±0.14}} & \cellcolor[HTML]{1F9F88}{\textcolor[HTML]{FFFFFF}{0.56±0.44}} & \cellcolor[HTML]{25848E}{\textcolor[HTML]{FFFFFF}{0.45±0.1}} & \cellcolor[HTML]{443B84}{\textcolor[HTML]{FFFFFF}{0.17±0.1}} \\ 
 & 5\% & \cellcolor[HTML]{D5E21A}{\textcolor[HTML]{000000}{0.94±0.016}} & \cellcolor[HTML]{8FD644}{\textcolor[HTML]{000000}{0.83±0.097}} & \cellcolor[HTML]{22A785}{\textcolor[HTML]{FFFFFF}{0.6±0.096}} & \cellcolor[HTML]{423F85}{\textcolor[HTML]{FFFFFF}{0.19±0.051}} \\ 
 & 12\% & \cellcolor[HTML]{C7E020}{\textcolor[HTML]{000000}{0.92±0.018}} & \cellcolor[HTML]{38B977}{\textcolor[HTML]{000000}{0.67±0.18}} & \cellcolor[HTML]{55C667}{\textcolor[HTML]{000000}{0.74±0.12}} & \cellcolor[HTML]{2E6E8E}{\textcolor[HTML]{FFFFFF}{0.36±0.085}} \\ 
 & 24\% & \cellcolor[HTML]{B1DD2E}{\textcolor[HTML]{000000}{0.89±0.051}} & \cellcolor[HTML]{74D055}{\textcolor[HTML]{000000}{0.79±0.002}} & \cellcolor[HTML]{21A585}{\textcolor[HTML]{FFFFFF}{0.59±0.027}} & \cellcolor[HTML]{2C718E}{\textcolor[HTML]{FFFFFF}{0.37±0.1}} \\ 
\midrule\addlinespace[2.5pt]
\multirow{4}{*}{Classifier (ST-GCN)} &1.5\% & \cellcolor[HTML]{ECE51B}{\textcolor[HTML]{000000}{0.97±0.011}} & \cellcolor[HTML]{B2DD2D}{\textcolor[HTML]{000000}{0.89±0.014}} & \cellcolor[HTML]{24AA83}{\textcolor[HTML]{FFFFFF}{0.61±0.0023}} & \cellcolor[HTML]{355F8D}{\textcolor[HTML]{FFFFFF}{0.3±0.007}} \\ 
&5\% & \cellcolor[HTML]{F3E61E}{\textcolor[HTML]{000000}{0.98±0.0077}} & \cellcolor[HTML]{C6E021}{\textcolor[HTML]{000000}{0.92±0.0038}} & \cellcolor[HTML]{73D056}{\textcolor[HTML]{000000}{0.79±0.016}} & \cellcolor[HTML]{218F8D}{\textcolor[HTML]{FFFFFF}{0.5±0.0032}} \\ 
&12\% & \cellcolor[HTML]{F9E622}{\textcolor[HTML]{000000}{0.99±0.0049}} & \cellcolor[HTML]{E3E418}{\textcolor[HTML]{000000}{0.96±0.0056}} & \cellcolor[HTML]{AEDC30}{\textcolor[HTML]{000000}{0.88±0.013}} & \cellcolor[HTML]{23A983}{\textcolor[HTML]{FFFFFF}{0.6±0.0075}} \\ 
&24\% & \cellcolor[HTML]{F6E620}{\textcolor[HTML]{000000}{0.99±0.0041}} & \cellcolor[HTML]{F1E51D}{\textcolor[HTML]{000000}{0.98±0.0016}} & \cellcolor[HTML]{C7E020}{\textcolor[HTML]{000000}{0.92±0.0035}} & \cellcolor[HTML]{47C16E}{\textcolor[HTML]{000000}{0.71±0.003}} \\ 
\bottomrule
\end{tabular}}
\caption{Architecture performance on \textbf{synthetic data}. The standardized area under the precision-recall curve of the two models for different levels of frequency for the rare behavior (rows, column 2) and similarity between the frequent and rare behaviors (columns 3-7). Performance ranges from poor (0.17, dark blues) to excellent (0.99, yellows). ST-GCN classifiers are superior across the board. When behavior is distinct (column 3) anomaly detection yields decent results with no labeling effort invested, however, performance degrades quickly with similarity.}
\label{table:simulated_aucs}
\end{table}

ST-GCN-based classifiers showed the best performance across the entire range of behavioral frequencies and behavioral similarities (see \Cref{figure:sim_prcs} and \Cref{table:simulated_aucs}). 
Surprisingly, if the behavior of interest is quite distinct (behavior SD = 0.5), even high rarity levels (translating into high data imbalance) only mildly affect performance (-0.02 standardized AuPRC for the lowest overlap level). 
A linear model predicting the standardized AuPRC as a function of the baseline behavioral frequency (rarity), behavioral similarity, and architecture found quite intuitive results. 
The unsupervised architecture was worse than the supervised one (coefficient = -0.19, $p<5.8e-13$).
The similarity between behaviors (behavior SD) hurts performance (coefficient =-0.15, $p < 0.0001$). The baseline rarity by itself had no significant effect, however, it positively interacts with behavior similarity (coefficient = 0.27, p $<$ 0.0001) which means that when a behavior is more frequent, behavior similarity has less effect on performance.
The linear model explained a substantial part of the variance in the data ($R^2=0.81$).

All in all these results establish that, given sufficient data, graph classifiers deliver high performance even under extreme imbalances. Additionally, they highlight that while rarity may not be an issue by itself when dealing with fine-grained behaviors it hurts performance considerably.
However, the question that remains is how we find the labeled instances of rare behaviors to train these highly performant classifiers.
This motivated us to find a way to quickly obtain and annotate rare behaviors using the anomaly detector which, though less performant, requires no labeled samples.

\subsubsection{Biological datasets} 
\begin{figure}[htbp] 
  \centering
    \includegraphics[scale=0.6]{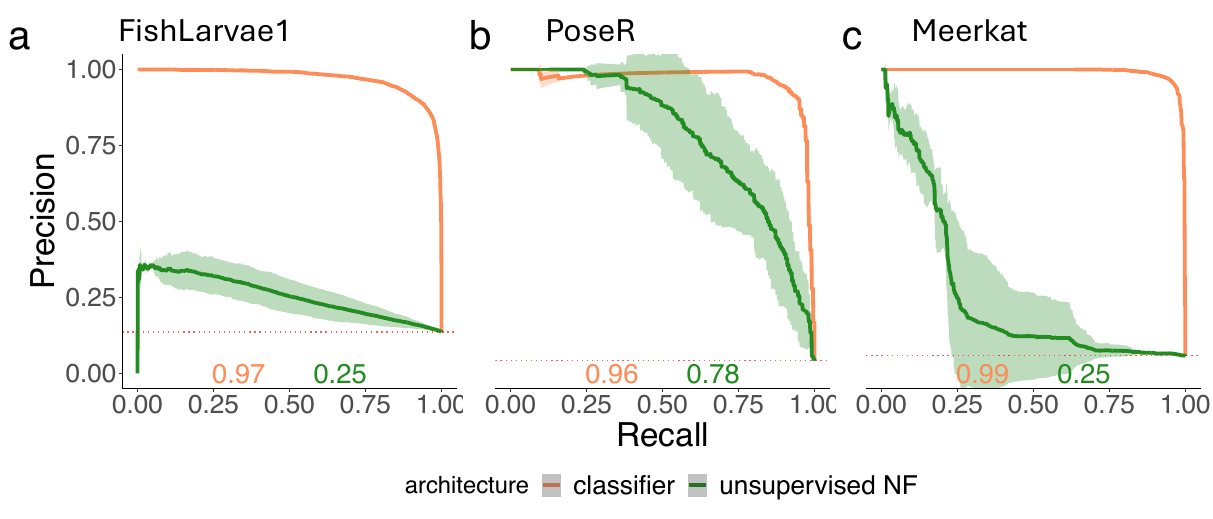}
    \caption{Architecture performance on \textbf{biological data}. Precision-Recall curves for each of the experimental datasets (panels a-c). ST-GCN classifiers (orange) dominate performance. Unsupervised Normalizing Flows (green) are not competitive on their own (except for panel b).}
    \label{figure:real_prcs}
\end{figure}

In all biological datasets, like the synthetic data, ST-GCNs show superior results (\Cref{figure:real_prcs}) despite the high data imbalance. 
Unsupervised STG-NFs, for 2 out of 3 datasets, show drastically lower performance. 
It has been previously shown that STG-NF is adversly affected when the train set has a high percentage of abnormal samples \cite{hirschorn2023normalizing}. 
During some preliminary experimentation, we found this to be partially, though not entirely, the case here.
Integrating insights from the synthetic regime, this reduced performance in the unsupervised approach implies a higher degree of similarity between rare and common behaviors.

{\small
\bibliographystyle{ieee_fullname}
\bibliography{supplementary}
}